# Accurate background velocity model building method based on iterative deep learning in sparse transform domain

Guoxin Chen[1]

1 Ocean College, Zhejiang University, Zhoushan, 316021, China.

Corresponding author: Guoxin Chen    Email: zjucgx@zju.edu.cn

# SUMMARY

Whether it is oil and gas exploration or geological science research, it is necessary to accurately grasp the structural information of underground media. Full waveform inversion is currently the most popular seismic wave inversion method, but it is highly dependent on a high-quality initial model. Artificial intelligence algorithm deep learning is completely data-driven and can get rid of the dependence on the initial model. However, the prediction accuracy of deep learning algorithms depends on the scale and diversity of training data sets. How to improve the prediction accuracy of deep learning without increasing the size of the training set while also improving computing efficiency is a worthy issue to study. In this paper, an iterative deep learning algorithm in the sparse transform domain is proposed based on the characteristics of deep learning: first, based on the computational efficiency and the effect of sparse transform, the cosine transform is selected as the sparse transform method, and the seismic data and the corresponding velocity model are cosine transformed to obtain their corresponding sparse expressions, which are then used as the input data and corresponding label data for deep learning; then we give an iterative deep learning algorithm in the cosine transform domain, that is, after obtaining the seismic data residuals and velocity model residuals of the previous round of test results, they are used again as new input data and label data, and re-trained in the cosine domain to obtain a new network, and the prediction results of the previous round are corrected, and then the cycle is repeated until the termination condition is reached. The algorithm effect was verified on the SEG/EAGE salt model and the seabed sulfide physical model site data.

**Key words:** Iterative deep learning; Sparse transform; Cosine transform ; Velocity model building; SCU-Net.

# Introduction

The velocity and path of seismic waves generated by artificially excited seismic sources will change in different media. The principle of seismic wave exploration is to infer the underground structure by analyzing the propagation characteristics of the seismic waves received and transmitted back to the surface. The application of seismic wave exploration covers hydrocarbons exploration, geological disaster monitoring, engineering geophysical exploration, etc. The velocity information of the underground medium in seismic wave exploration is the most critical geophysical parameter. The more commonly used velocity model building methods are migration velocity analysis (Symes, 2008), travel time tomography (Hole, 1992) and full waveform inversion (Tarantola, A., 1986). Compared with the first two, full waveform inversion uses complete seismic waveform information, including the amplitude, frequency and phase of the wave, to reconstruct the underground structure with high resolution. Full waveform inversion has been deeply studied and applied in geological science research (Górszczyk et al. 2021; Yang et al., 2024), oil and gas exploration (Routh et al. 2017; Chen et al., 2019, 2022), and shallow surface engineering geophysical exploration (Zhang et al., 2024). However, in order to make it truly practical, full waveform still has many challenges and technical bottlenecks to solve. The strong nonlinear relationship between seismic data and velocity determines that full waveform inversion is a strongly nonlinear inverse problem. To obtain the global optimal solution, the initial velocity model cannot have a significant deviation from the true model. This is a condition that is difficult to meet in actual production.

In order to solve the dependence of full waveform inversion on the initial model, many scholars have proposed many different solutions from improving the convexity of the objective function (Shin and Cha, 2008; Chen et al., 2017, 2018, 2019, 2020; Engquist et al., 2016; Métivier et al., 2016) and changing the solution algorithm (Wu and Zheng, 2014; Innanen, 2015). With the rapid improvement of computer software and hardware technology, data-driven artificial intelligence inversion algorithms have received extensive attention and research, among which deep learning has become a new model for solving the problem of local minima problem of the full waveform inversion. Deep learning establishes a nonlinear mapping from seismic data to parameter models through multi-layer neural networks (Araya-Polo et al., 2018; Yang and Ma, 2019, 2023; Sun et al., 2020). In order to solve the problem of the dependence of the inversion accuracy and generalization ability of supervised deep learning on training data, some scholars apply self-supervised or unsupervised deep learning to the inversion of underground medium parameters (Sen and Dhara, 2022; Liu et al., 2023), or add seismic wave field constraints to neural networks (Song et al., 2021; Dhara and Sen, 2023), or use neural networks to optimize full waveform inversion parameters (Sun and Alkhalifah, 2020). Using deep learning to build a high-quality initial model for conventional full waveform inversion is also a relatively intuitive implementation strategy of digital-analog joint drive (Muller et al., 2023; Chen et al., 2024).

Using deep learning to provide a high-quality initial model for full waveform inversion is a new way to solve the problem of full waveform inversion's dependence on the initial model. However, the prediction accuracy and generalization ability of deep learning are heavily dependent on the scale and quality of training data. Therefore, how to improve the prediction accuracy and generalization ability of deep learning without significantly increasing the scale of training data sets is crucial to the practical application of deep learning. This paper studies the above problems. We use the SCU-Net network that combines U-net and Transformer to improve the data feature extraction ability of the neural network, and then propose an iterative deep learning implementation strategy in the cosine transform domain to improve the prediction accuracy and generalization ability of deep learning. Finally, the SEG/EAGE Salt model and the physical model inversion effect of seabed sulfide verify the algorithm proposed in the article. Finally, in the discussion section, we discuss the limitations of this method and future research directions.

**Theory and method**

### Iterative deep learning algorithm in cosine transform domain

To ensure the quality of the initial model constructed by deep learning, it is necessary to improve the prediction accuracy and generalization ability of deep learning. The reason why deep learning cannot obtain accurate prediction results is mainly because the data features in the training data set do not cover the test data features. Therefore, effectively improving the coverage of the training data set for the inversion target data features, comprehensively and accurately extracting the data features of the training data, and strengthening the re-learning ability of the data features not covered by the training data are the key points to improve the generalization ability of deep learning. However, in seismic exploration, the geological structure characteristics of the exploration area are ever-changing, the acquisition equipment and environment are also different, and the corresponding seismic responses are also completely different. In addition, the amplitudes of different data features or their proportions in the total features are not the same. For example, the amplitudes of the first arrival wave and the deep reflection wave in seismic data are orders of magnitude different, and the proportions of the macroscopic reflection interface and the small-scale scatterers in the parameter model features in the parameter model are also quite different. This significant difference in amplitude or proportion will cause some data features with small amplitudes or proportions to be masked or ignored in network training, resulting in the loss of data features. How to ensure that all data features of the training data set can be extracted and participate in the optimization training of network parameters is also an important research topic to ensure the generalization ability of deep learning. In order to solve the above two problems, we have given corresponding solutions.

**Improving the generalization ability of deep learning by implementing deep learning in the cosine transform domain**

Effectively improving the coverage of target data features in the training data set is the key to ensuring the generalization ability of deep learning. The main reason why it is difficult to integrate all data features into the training data set is that the data features are large and severely redundant in the time and space domain. Therefore, a certain sparse transformation can be adopted to sparsely express the seismic data and the corresponding parameter model, and use the transformation domain coefficients as new input data and label data. This can significantly reduce the scale of deep learning training data and improve the efficiency of the algorithm. More importantly, the data features are highly compressed, reducing their dimensions and redundancy, so that two data features with significant differences in the time and space domain are highly similar and overlapping in the transformation domain. When choosing a sparse transformation method, because it is necessary to process a large amount of training data in deep learning, in addition to considering the sparse expression effect of the transformation method, it is also necessary to take into account the difficulty of its operation and computational efficiency.

In terms of data sparse expression ability, the compression ability of two-dimensional discrete cosine transform for images and data has been verified in the field of signal processing. In terms of operation, two-dimensional discrete cosine transform has a mature and fast algorithm. Compared with other transformation methods, such as wavelet transform and curvelet transform, it has the advantages of low computational complexity and high computational efficiency. In view of this, this project uses two-dimensional discrete cosine transform for seismic data $u$, the velocity $V$ of the underground medium:

$$\begin{aligned} \tilde{W}_u &= \mathrm{R}(\mathrm{DCT}(u)), \\ \tilde{W}_V &= \mathrm{R}(\mathrm{DCT}(V)), \end{aligned} \quad (1)$$

Where DCT is the operator of the two-dimensional discrete cosine transform, and R is the truncation operator, which clips the two-dimensional cosine transform coefficients according to the set threshold and only retains the distribution area of the dominant coefficients. $\tilde{W}_u, \tilde{W}_V$ are the 2D cosine transform domain coefficients of $u, V$ after 2D discrete cosine transform and clipping.

In order to further improve the sparsity of the data and consider the bias of deep learning towards low-frequency features (reference), the seismic data and velocity model can be preprocessed before the sparse transformation. The purpose of preprocessing is to retain the macroscopic structural features of the seismic data and parameter model while removing the details of high frequency or high beam, thereby enhancing the complexity of the data in the time and space domain, and then enhancing the sparsity of the sparse transformation domain. For the velocity model, Gaussian smoothing can be used.

$$V_G(x,y) = \frac{1}{2\pi\sigma^2} e^{-\frac{x^2+y^2}{2\sigma^2}} \quad (2)$$

Among them, $V_G$ represents the velocity model after Gaussian smoothing, and $\sigma$ represents the standard deviation of Gaussian distribution. For seismic data, it is an intuitive approach to use low-pass filtering in the frequency domain. However, in actual situations, the low-frequency components of seismic data are often missing, and are also accompanied by strong random noise or coherent noise interference, so low-pass filtering is not feasible. How to extract the main macro information of seismic data, or skeleton information, the window average envelope is a simple low-pass filtering method, but the difference between it and conventional low-pass filtering is that by solving the envelope of seismic data, the missing low-frequency components can be restored, while the introduction of window average retains the low-frequency components of seismic data, thereby extracting the skeleton information of seismic data. Here we choose the window average function instead of the conventional linear filtering method based on the renormalization strategy (Wilson and Kogut, 1974):

$$e^2 = u^2 + u_h^2$$
$$e_L^2(t) = \frac{1}{L}\int_{-L/2}^{L/2} dt' W(t-t')\left[e^2(t')\right] \quad (3)$$

$y$ is the synthetic data, $y_H$ is the imaginary part of the corresponding analytical signal obtained after Hilbert transform. $W(t)$ is the Gaussian window function, and $L$ is the window width. Compared with the linear filtering method, the window average function has more physical meaning: by using the window average to widen the width of the envelope energy flow, the smooth Gaussian window can avoid the Gibbs effect. Correspondingly, the two-dimensional discrete cosine transform of the preprocessed seismic data and velocity model：

$$\tilde{W}_{e_L} = \mathrm{R}(\mathrm{DCT}(e_L)),$$
$$\tilde{W}_{V_G} = \mathrm{R}(\mathrm{DCT}(V_G)). \quad (4)$$

Figures 1 and 2 show the single shot seismic records obtained by forward modeling on the SEG/EAGE salt dome model, the transform domain coefficients of the salt dome model after cosine transform, and the transform domain coefficients after preprocessing. It can be seen that preprocessing has a significant effect on improving the sparsity of the cosine transform domain coefficients. For the sake of ease of explanation, the following text defaults to using the preprocessed seismic data and velocity model for cosine transform.

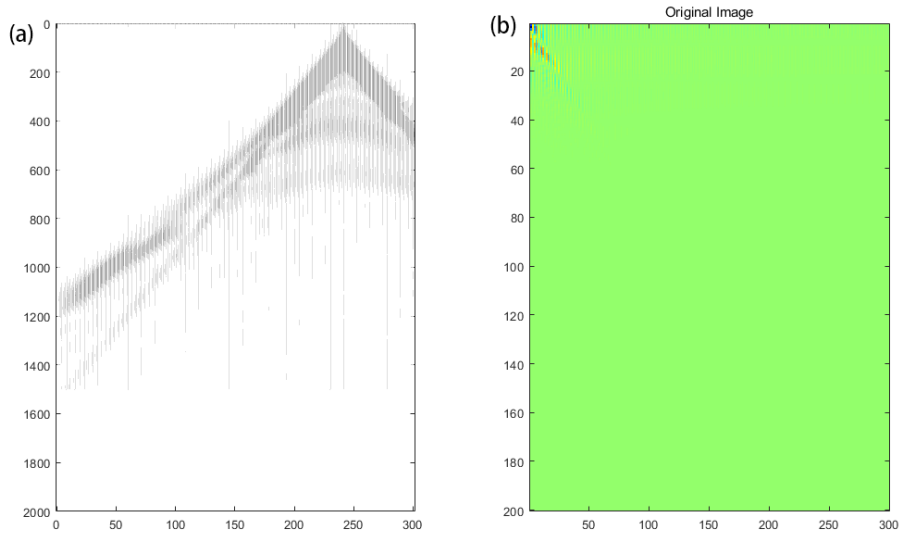

Figure.1 (a) Smoothed envelope data of a single shot record obtained by forward modeling on the SEG/EAGE Salt model, where the window width is 51; (b) Cosine transform result of a single shot record.

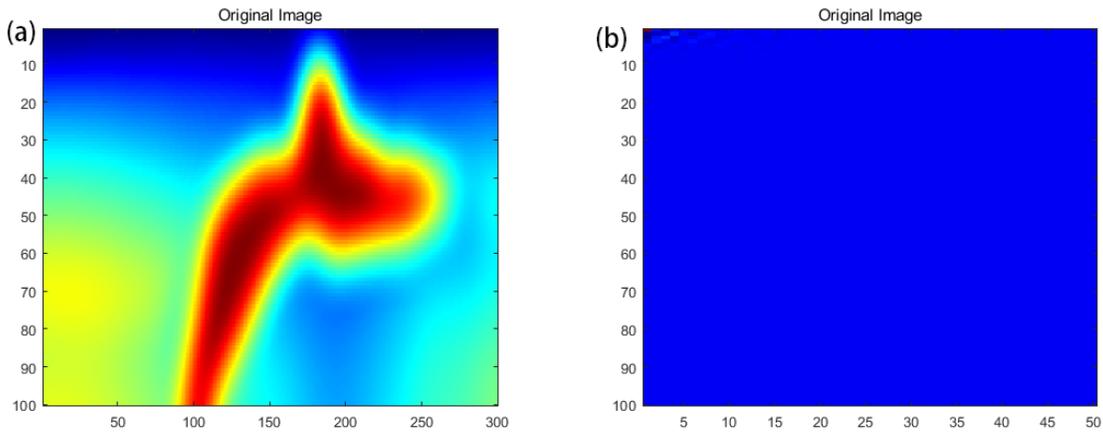

Figure.2 (a) Salt model after smoothing; (b) cosine transform of the Salt model

Choosing a suitable neural network is crucial to improving the prediction accuracy of deep learning. U-net is a widely used neural network with an encoding-decoding "U" structure. Although U-net has achieved good results in many applications, the network's ability to learn global semantic information needs to be improved. The Swin-Transformer (SwinT) model borrows the hierarchical structure of convolutional neural networks, obtains self-attention in non-overlapping local windows through sliding windows, and significantly reduces computational costs through cross-window connections. The Swin-Conv module combines the advantages of SwinT and residual convolution (RConv), and can be used as the main component of the U-net framework for up and down sampling. Chen et al (2024a,b) organically integrated the two complementary network architecture designs of U-net and Swin-Conv, and proposed a neural network structure with higher data feature extraction capabilities, namely SCU-net (Swin Conv U-net). The nonlinear mapping based on SCU-net in the cosine transform domain can be written as:

$$\tilde{W}_{V_P} = Net_{\tilde{W}_{V_P}}\left(W_{c_L}; \Theta_{W_{V_P}}\right), \tag{5}$$

Where $\tilde{W}$ and $W$ represent the cosine transform domain coefficients of the predicted value and the true value respectively, and $\Theta$ represents the optimization parameter of the network. In the cosine transform domain, we use the L1 norm to enhance the sparsity of the objective function. In addition, in order to ensure that the predicted results and the true results remain similar in structure, we introduce the structural similarity constraint SSIM into the objective function:

$$SSIM(\alpha, \beta) = \frac{(2\mu_\alpha\mu_\beta + c_1)(2\sigma_{\alpha\beta} + c_2)}{(\mu_\alpha^2 + \mu_\beta^2 + c_1)(\sigma_\alpha^2 + \sigma_\beta^2 + c_2)} \tag{6}$$

Where $\mu_\alpha, \mu_\beta, \sigma_\alpha, \sigma_\beta, \sigma_{\alpha\beta}$ is the mean, variance and covariance of $\alpha, \beta$. $c_1, c_2$ is the regularization parameter. After training, the neural network can be used to predict model parameters. The prediction results are restored to their original dimensions after zero padding, and then the underground medium parameter model is obtained through inverse cosine transformation.

**Improving generalization capabilities through multiple rounds of neural network iterative training**

Whether the data features in the training data set are fully and accurately extracted, and whether the neural network has the ability to re-learn data features not covered by the training data set are also important factors related to the generalization ability of deep learning. In response to the above problems, this project proposes an implementation strategy of multiple rounds of neural network serial iterations. The specific operation is as follows: For the first round of neural networks, the input data is seismic data, and the label data is the parameter model. After the training is completed, the trained neural network is used to test the training data and obtain the corresponding prediction results. Then, forward simulation is performed based on the prediction results and the real model, the corresponding seismic data is obtained and subtracted to obtain the seismic wave residual envelope $\delta e_L$. The prediction results and the real results are subtracted to obtain the residual of the model parameters $\delta V_G$. Then for the second and subsequent training rounds, the input data is the residual envelope $\delta e_L$, and the label data is the difference $\delta V_G$ between the predicted parameter model result and the true model. The nonlinear mapping in the cosine transform domain is:

$$\tilde{W}_{\delta V_G} = Net_{\tilde{W}_{\delta V_G}} \left( W_{\delta e_L}; \Theta_{W_{\delta V_G}} \right) \tag{7}$$

Similar to the first round, the objective function used in subsequent rounds of iterations is:

$$E_{\delta V_G} = \frac{1}{N} \sum_x \sum_z \left| \tilde{W}_{\delta V_G}(x,z) - W_{\delta V_G}(x,z) \right| + \frac{\left( 2\mu_{\tilde{W}_{\delta V_G}} \mu_{W_{\delta V_G}} + c_1 \right)\left( 2\sigma_{\tilde{W}_{\delta V_G} W_{\delta V_G}} + c_2 \right)}{\left( \mu^2_{\tilde{W}_{\delta V_G}} + \mu^2_{W_{\delta V_G}} + c_1 \right)\left( \sigma^2_{\tilde{W}_{\delta V_G}} + \sigma^2_{W_{\delta V_G}} + c_2 \right)} \tag{8}$$

In the second and subsequent rounds, the results of this round are used to supplement and update the prediction results of the previous round to obtain a parameter model that is closer to the actual results. Then, the new residual wave field and parameter model residual are calculated and put into the next round of network training. The number of iterations can be determined according to the correction amount of the parameter model. If the correction amount of the parameter model is close to zero or less than the set threshold, the iteration can be stopped. By setting up multiple rounds of neural networks and passing the prediction error of the previous round to the neural network of the following rounds, it is ensured that the data features in the training data set can be extracted and participate in the network training. More importantly, it also gives the neural network the ability to re-learn data features that are not covered by the training data set, thereby improving the generalization ability of deep learning. Figure 3 shows the implementation process of iterative deep learning in the cosine transform domain using the SCU-Net:

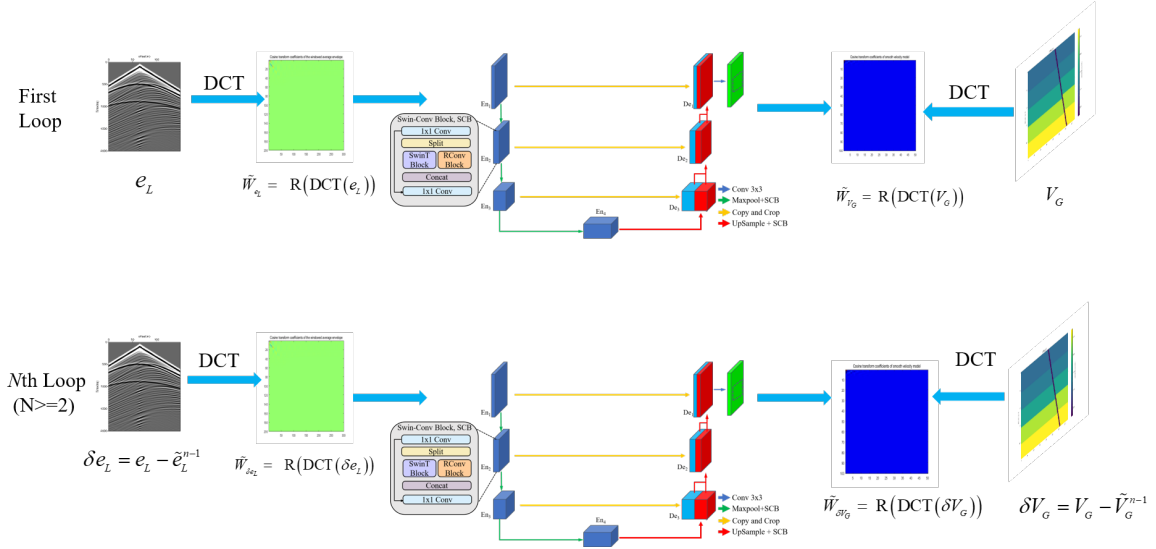

Figure.3 the implementation process of iterative deep learning in the cosine transform domain using the SCU-Net

**Numerical experiments on 2-D SEG/EAGE Salt model**

The SEG/EAGE salt model (Figure 4a) is a typical salt structure. The difficulty of inversion lies in its huge salt dome. To obtain the internal velocity of the salt dome, the upper and bottom boundaries require extremely low-frequency seismic data. To obtain the precise subsalt structure, it is necessary to break through the barrier of the salt dome, which is a strong scattering medium. Large offset seismic data is often required, and these data are often missing in the field data. Figure 4c is the inversion result obtained by FWI on the linear gradient initial model (Figure 4b). It can be seen that it is difficult to accurately reconstruct the subsalt structure of the salt dome without low-frequency seismic data and large offset data. Assuming that the previous work has confirmed that the underground medium contains salt domes, but the velocity inside the salt dome and the location of the salt dome body are uncertain, we modified the SEG/EAGE salt dome model to change the location, shape, velocity, and upper and bottom boundary positions of the salt dome body.

In order to verify the performance of the algorithm under limited samples, we only designed 200 training models. Figure 5 shows several typical representatives of the samples.

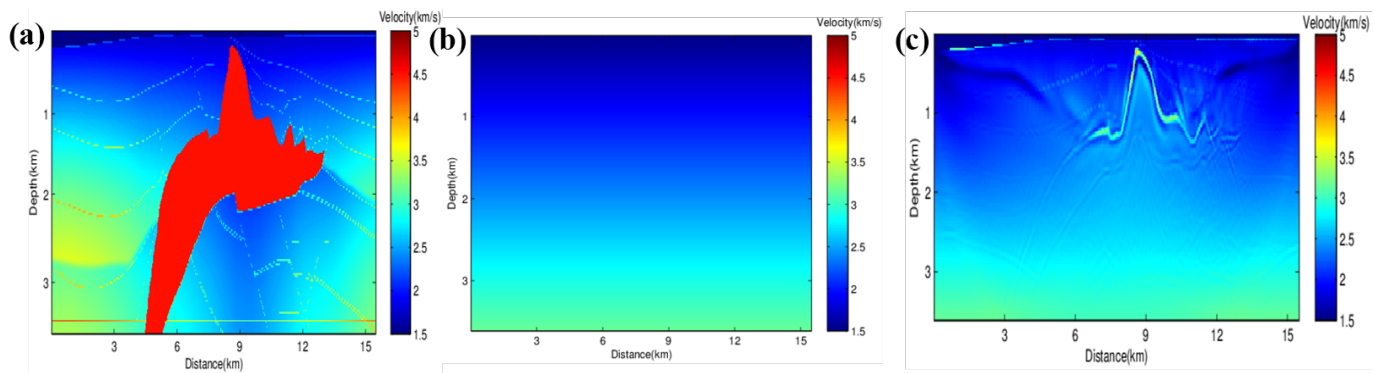

Figure.4 (a) SEG/EAGE salt dome model; (b) initial model; (c) conventional full waveform inversion results.

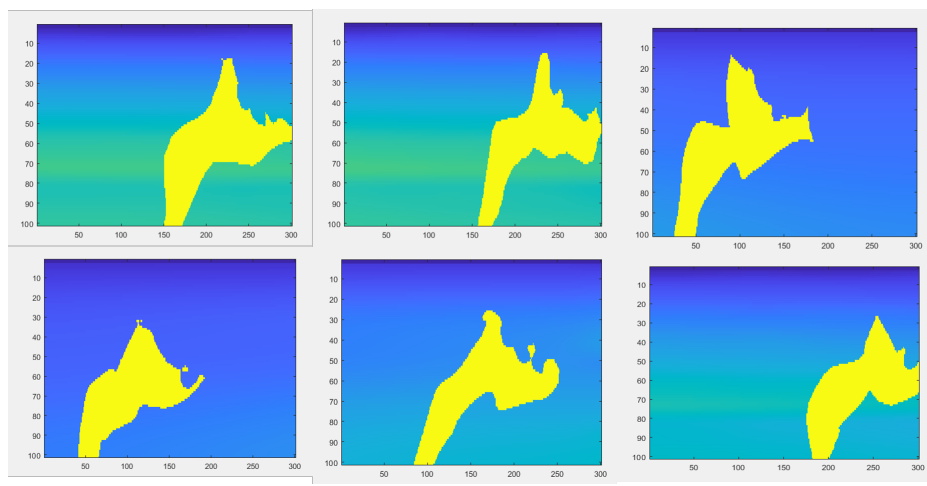

Figure.5 Some velocity model training samples

We used finite differences to perform forward modeling on the velocity model of the training dataset, and each set of models corresponded to 5 shots of seismic data as input data. The test dataset was the SEG/EAGE salt model, and the same observation system was adopted for the training data. The Relu activation function was used, and then the Adam optimization algorithm was used for gradient calculation. We first give the inversion results obtained by the conventional processing method. Figure 6 shows the prediction results after 500 iterations, from which it can be seen that there is a significant difference between the prediction results and the actual results. In the algorithm proposed in this paper, for seismic data, the envelope with a window length of 151 is calculated using Equation 3. For the velocity model, the dominant coefficient is retained after the two-dimensional discrete cosine transform. After 500 iterations, the inversion result is obtained, and the corresponding inversion result is obtained after the inversion result is filled with zero and the cosine inverse transform is performed (Figure 7b). Comparing the inversion results with the real model, it can be seen that the positioning and velocity of the salt dome have a good match with the real model. Then a second round of iterations is carried out(Figure 7c). Comparing these rounds of results, it can be seen that by conducting targeted training on the residuals, the errors in the previous round of prediction results can be corrected in a targeted manner. In order to give a detailed comparison, we give a comparison chart of the velocity profile curves of different inversion results at a horizontal position of 2km(Figure 7d).

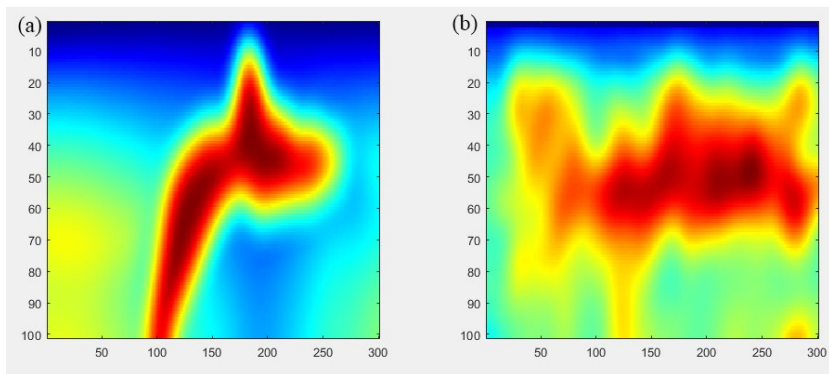

Figure.6 The smoothed result of the real Salt model; the prediction results obtained using the SCU-Net network.

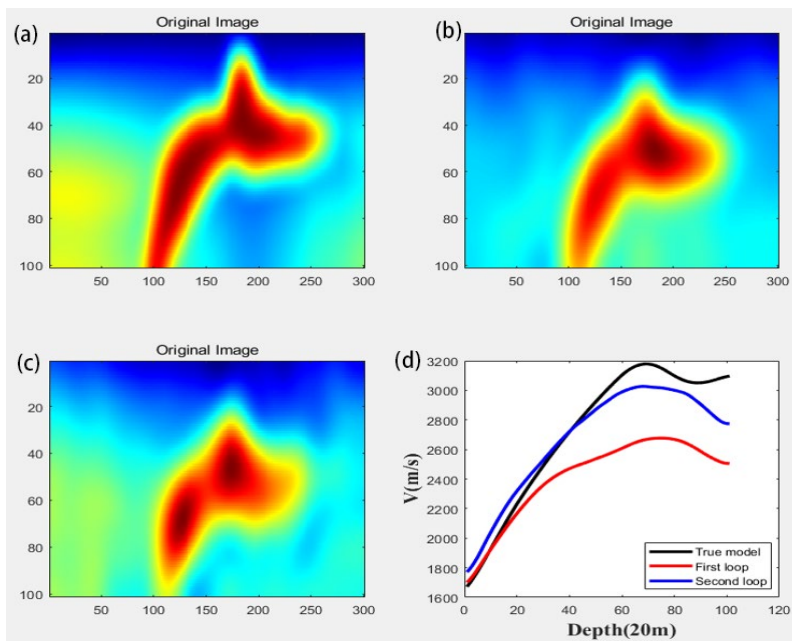

Figure.7 (a) The smoothed result of the real Salt model; (b) The first round prediction result using the cosine domain SCU-Net; The second round prediction result using the cosine domain SCU-Net; Single-track velocity comparison of prediction results in different rounds.

**Numerical experiments on Seafloor sulfide physical model**

Seafloor sulfides are rich in metal mineral resources, but they are usually located at a depth of several thousand meters on the seafloor, and their length and width do not exceed one kilometer. Therefore, conventional towed cable seismic exploration is too far from the target body, resulting in weak energy of the related seismic reflection waves received, and OBS is not suitable for seafloor sulfide exploration due to its low resolution. Near-bottom seismic is a more suitable observation system, which tows the source/receiver array in the seawater near the seafloor or fixes it vertically on the seafloor for detection, which can compress the radius of the first Fresnel zone and improve the lateral resolution. In order to verify the effect of the near-bottom observation system on seafloor sulfide exploration, we completed the physical model (Figure 8b) design according to the two-dimensional geological structure (Figure 8a). The simplified model mainly includes surrounding rock (basalt), volcanic rock (massive sulfide, stockwork sulfide and alteration zone) and fractures. The scale ratio of the model is 1:1000, the velocity ratio is 1:2, and the actual size of the model finally established is 800mm*600mm*300mm, and the simulated actual work area range is 800m*600m*300m.

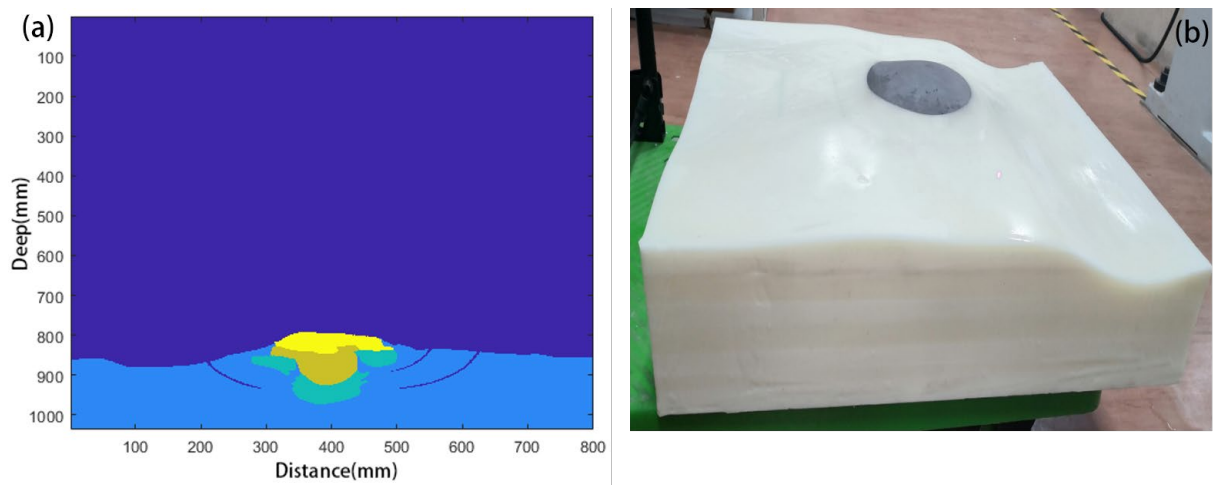

Figure.8 (a) 2D numerical model; (b) physical model

In order to verify the effect of the actual observation system, we use the same source and observation system as the physical model in the numerical model. The vertical seismic data observation test is carried out using an 80KHz sound source.

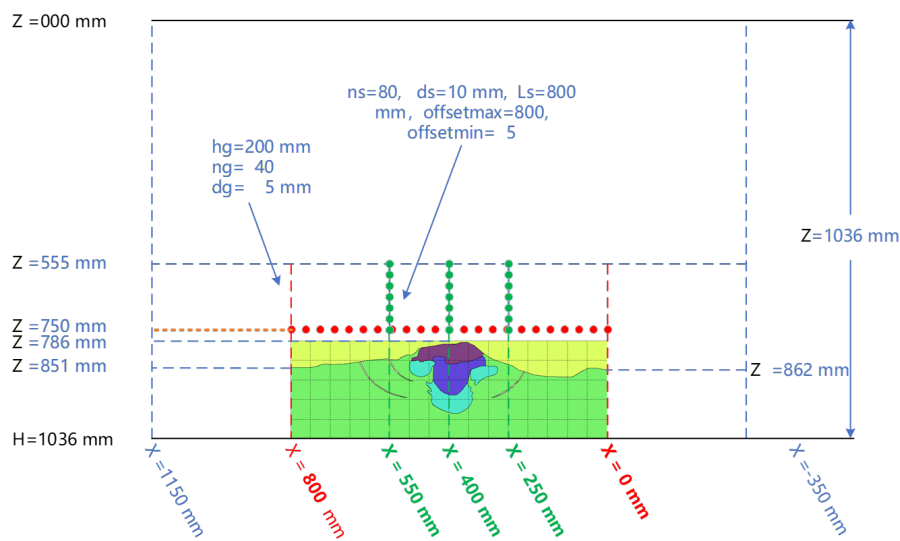

Figure.9 Observation system

The observation method is shown in Figure 9, where the green dotted line represents the receiving array, with a total of 3 arrays, vcs01 located at X=400mm, vcs02 located at X=550mm, and vcs03 located at X=250mm. Each array has 20 channels, and the channel spacing is 10mm. The top receiving point depth position is 555 mm, and the bottom receiving point depth position is 750 mm. The red dots are excitation points, with a total of 85, and the shot spacing is 10mm. The first shot excitation position is 0 mm, the last shot excitation position is 820 mm, and the excitation spacing is 10mm; the number of time sampling points is 10000, the time sampling interval is 1e-6 s, and the record length is 10ms. The following Figure 10 shows the common shot point gather data of the first shot obtained by forward simulation on the numerical model. Figure 11 shows some velocity model samples in the training set. Figure.12 (a) shows the smoothing of the true physical model and Figure.12 (b) shows the first round prediction result using the cosine domain SCU-Net . Figure.12 (c) shows difference between the predicted result and the real model. Figure 12(d) shows the comparison of the velocity profile between the predicted result and the real model. It can be seen that the proposed method can also achieve good results on actual field data.

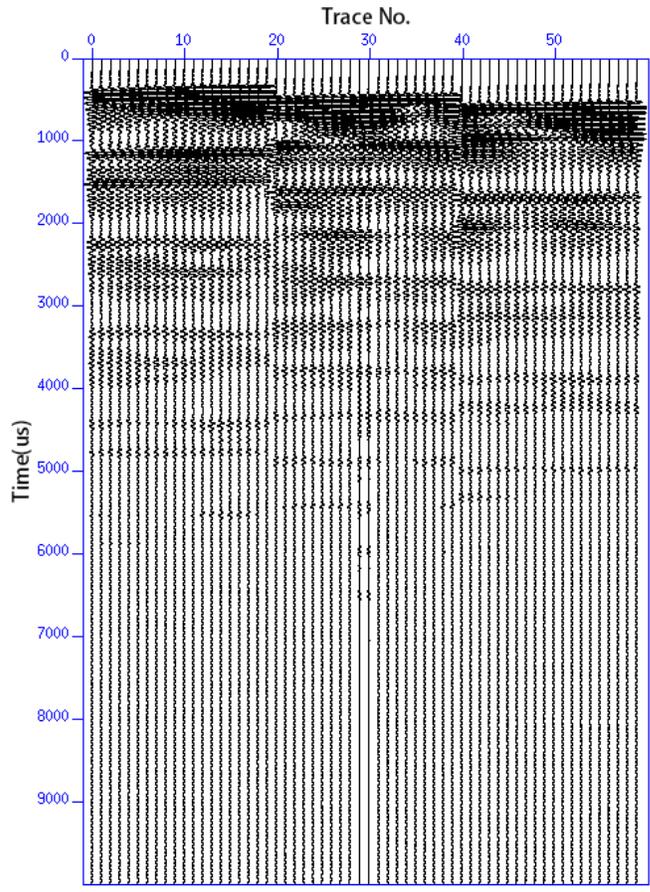

Figure.10 Common shot point gather data of the first shot.

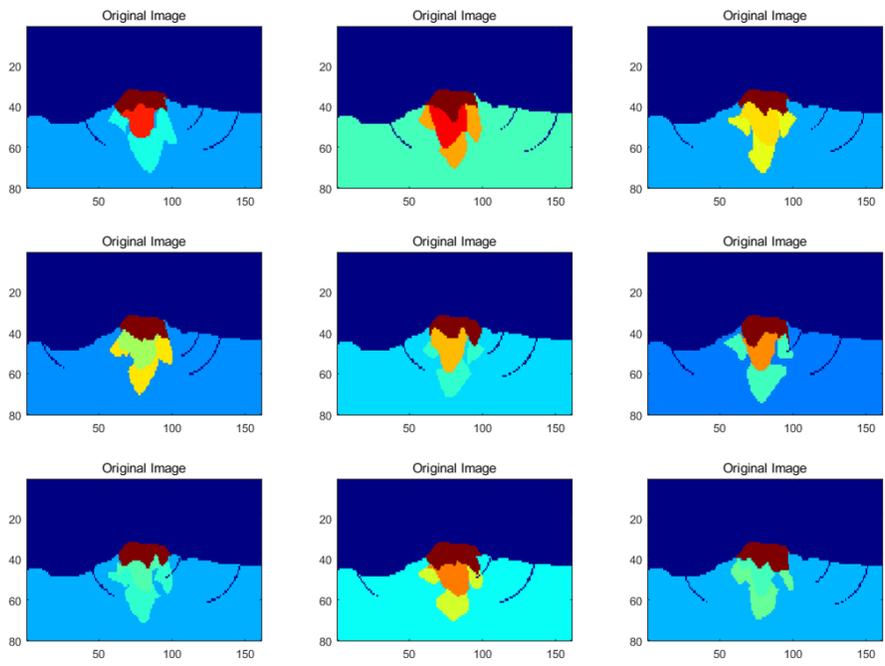

Figure.11 Some training samples

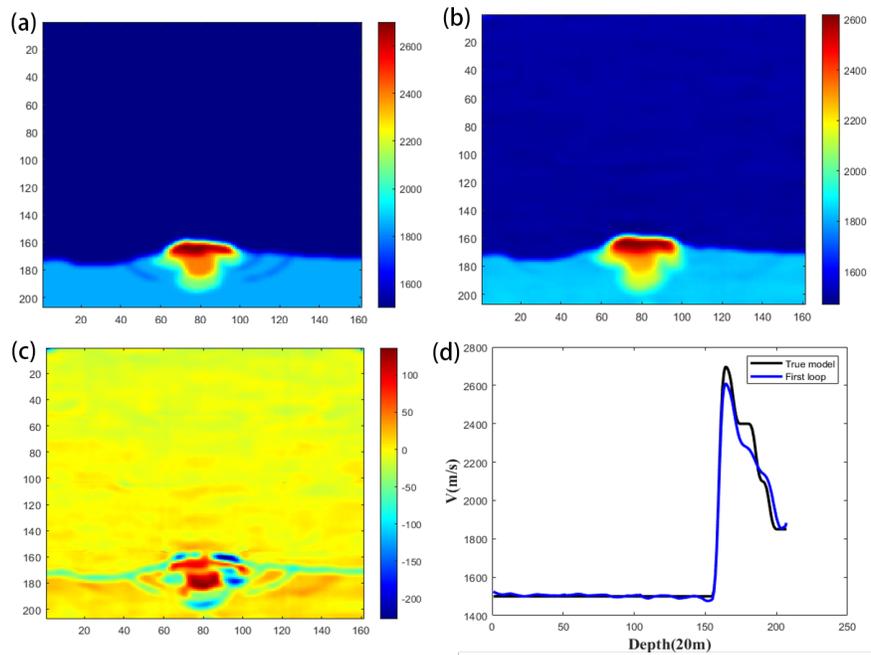

Figure.12 (a) The result after smoothing of the real physical model;(b) The first round prediction result using the cosine domain SCU-Net; (c) The difference between the predicted result and the real model;(d) Comparison of the velocity profile between the predicted result and the real model.

**Discussion and Conclusion**

Sparse transformation greatly reduces the dimension of the training data set, which not only improves the computational efficiency, but more importantly reduces the dimension of the solution space, thereby greatly improving the convergence of the network and correspondingly improving the prediction accuracy of the algorithm. On the other hand, the gap between data features with significant differences in the time and space domains is significantly compressed in the sparse transformation domain, thereby solving the problem of significant decrease in prediction accuracy caused by differences in the data features of the training set and the test set. How to express data features more sparsely is an important research direction. This is also our future research direction. Solving the window average envelope of seismic data and smoothing the velocity model is also a simple strategy to reduce the complexity of data features. This strategy can promote the sparse expression of data, and on the other hand, it also helps to improve the algorithm's anti-noise ability. This data preprocessing method also plays a very important role in improving the accuracy of the algorithm. How to express data features more sparsely is an important research direction. This is also our key research direction in the future. The window average envelope of seismic data and smoothing of the velocity model are also simple strategies to reduce the complexity of data features. This strategy can promote the sparse expression of data, and on the other hand, it also helps to improve the algorithm's anti-noise ability.This data preprocessing method also plays a very important role in improving the accuracy of the algorithm. The multi-round iteration strategy is equivalent to giving deep learning an opportunity to correct errors, thereby relearning the data features that were not learned in the previous round, thereby improving the ability to extract data features and further improving the prediction accuracy of the algorithm. The experimental results of the EAGE/SEG Salt model and the seabed sulfide physical model verified the effectiveness of the algorithm. The algorithm's anti-noise ability and its effect on more field data are important research directions for our future.

**Acknowledgments**

This work was financially supported by the Zhejiang Provincial Natural Science Foundation(Grant LY23D040001), Open Research Fund of Key Laboratory of Engineering Geophysical Prospecting and Detection of Chinese Geophysical Society (CJ2021GB01), Open Research Fund of Changjiang River Scientific Research Institute(CKWV20221011/KY), National Natural Science Foundation of China (Grant 41904100), HPC Center of ZJU (ZHOUSHAN CAMPUS). The authors report no conflicts of interest.